\documentstyle[12pt]{article}
\input tcilatex
\QQQ{Language}{
American English
}

\begin{document}

\title{{\large {\bf QUALITATIVE FEATURE OF THE LOW-LYING SPECTRUM OF INTRASHELL
STATES OF 4-VALENCE-ELECTRON ATOMS DERIVED FROM SYMMETRY CONSIDERATION}}}
\author{{C.G.Bao} \\
{\small Department of Physics, Zhongshan University, Guangzhou,510275,PRC}}
\date{}
\maketitle

\vskip -0.1in ABSTRACT: Inherent nodal surfaces existing in the
wavefunctions of intrashell states of 4-valence-electron atoms have been
investigated. The decisive effect of these surfaces has been demonstrated,
the ordering of low-lying levels has been predicted, and a primary
classification scheme has been proposed

\vskip 0.2in



\newpage 

The important role of symmetry in nature is well known. A number of laws (or
constraints) based on symmetry have been discovered, they govern the
evolution of basic physical processes in nature [1-5]. In quantum mechanical
systems, owing to the invariance of the Hamiltonian with respect to
symmetric operations (rotation, space inversion, particle permutation,
etc.), the eigenstates are classified according to a set of quantum numbers
, these numbers specify the transformation property of the wavefunctions
under symmetric operations. Since an eigenstate must have a specific
transformation property, the distribution of wavefunctions in coordinate
space may be greatly constrained. It was found in [6-8] that the constraint
is embodied by the existence of a special type of nodal surfaces, the so
called inherent nodal surface (INS). In this paper, we shall demonstrate how
the INS affect the spatial distributions of the wavefunctions of 4-electron
atoms, and how the low-lying spectrum is thereby decisively affected.

Let us first study the origin of the INS\ in a general way. Let A\ be a
point in a multi-dimensional coordinate space. Evidently, A is associated
with a geometric configuration. Let $O_i$ be a combined symmetric operation.
In some cases, one may find a set of $O_i$ ($i=$1 to m) to leave A
invariant; i.e., $O_iA=A$ (For example, is the case of a 4-body system an
equilateral tetrahedron configuration (ETH) is invariant to the rotation
about a 2-fold axis together with two interchanges of the particles, refer
to Fig.1). Let $\Psi $ be an eigenstate. Let $\hat O_i$ be the operator
acting on $\Psi $ so that $\hat O_i\Psi (A)=\Psi (O_iA)$. When $A$ is
invariant to $O_i$ , we have 
\begin{equation}
\hat O_i\Psi (A)=\Psi (A).\hspace{1.0in}({\rm i=1}\text{ }{\rm to}\text{ }%
{\rm m})
\end{equation}
Owing to the inherent transformation property of $\Psi $, eq.(1) always can
be written in matrix forms (as we shall see). In this form we have m sets of
homogeneous linear algebra equations. These equations impose a very strong
constraint on $\Psi $. Since a set of homogeneous equations does not always
have non-zero solutions, in some cases $\Psi $ must be zero at $A.$ This is
the origin of the INS. In the follows we shall concern mainly the 4-electron
atomic systems, however the generalization to other quantum systems is
straight forward.

An important geometric configuration with the strongest geometric symmetry
for the 4-body systems is the equilateral tetrahedron (ETH). Let $p_{ij}$
denotes an interchange of i and j. Let k' be a 2-fold axis of an ETH so that
a rotation of the ETH\ about k' is equivalent to $p_{12}p_{34}.$ Let i' be
an axis originating from the nucleus vertical to k' and parallel to $%
\stackrel{\rightarrow }{r_2}-\stackrel{\rightarrow }{r_1}$, where $\stackrel{%
\rightarrow }{r_i}$ is the position vector of the i-th electron originating
from the nucleus. . Let $R_\delta ^n$ be an operator of rotation about the $%
n $-axis by $\delta $ (in degree), let $P$ be the operator of a space
inversion. One can prove that the ETH is invariant to 
\begin{equation}
O_a=p(1423)PR_{90\hspace{1.0in} }^{k^{\prime }}
\end{equation}
\begin{equation}
O_b=p_{12}PR_{180\hspace{1.0in} }^{i^{\prime }}
\end{equation}
\begin{equation}
O_c=p(243)R_{120\hspace{1.0in} }^{\stackrel{\wedge }{r_1}}
\end{equation}
where $p(1423)$ and $p(243)$ denote cyclic permutations, $\stackrel{\wedge }{%
r_1}$ is a unit vector along $\stackrel{\rightarrow }{r_1}$. There are other
operators leave the ETH also invariant, e.g., the $p(134)R_{120\hspace{1.0in}
}^{\stackrel{\wedge }{r_2}}$. However, since $R_{120}^{\stackrel{\wedge }{r_2%
}}=R_{-180}^{k^{\prime }}R_{120}^{\stackrel{\wedge }{r_1}}R_{180,}^{k^{%
\prime }}$ this operator does not introduce new constraints. Therefore the
above three are sufficient to specify the constraints arising from symmetry.

In the case of an atom with four valence electrons (the degrees of freedom
of the core is neglected) , an eigenstate can be in general expanded as

\begin{equation}
\Psi ={\sum_i}F_{Mi}^{L\Pi \lambda }(1234)\chi _i^{\stackrel{\sim }{\lambda }%
}
\end{equation}
Where, L is the total orbital angular momentum, $\Pi $ is the parity, $%
\lambda $ denotes a representation of the S$_4$ group, $i$ labels a basis
function of the $\lambda $-representation, $M$ is the Z-component of L. $%
F_{Mi}^{L\Pi \lambda }(1234)$ is a function of the spatial coordinates. $%
\stackrel{\sim }{\lambda }$ is the conjugate representation of $\lambda $ , $%
\chi _i^{\stackrel{\sim }{\lambda }}$ is a basis function of the $\stackrel{%
\sim }{\lambda }-$representation in spin space. It is well known that $%
\lambda $ is determined by the total spin S of the atom, we have $\lambda =$%
\{2,2\} if S=0, $\lambda =$\{$2,1,1$\} if S=1, or $\lambda =$\{1$^4$\} if
S=2.

From the i' and k' axis, one can define a body frame i'-j'-k'. In this frame
the spatial function can be expanded as 
\begin{equation}
F_{Mi}^{L\Pi \lambda }(1234)=\sum_QD_{QM}^L(-\gamma ,-\beta ,-\alpha
)F_{Qi}^{L\Pi \lambda }(1^{\prime }2^{\prime }3^{\prime }4^{\prime })
\end{equation}
where $D_{QM}^L$ is the Wigner function of rotation, $\alpha \beta \gamma $
are the Euler angles for the collective rotation, Q is the component of L
along the k' axis, (1234) and (1'2'3'4') denote the coordinates relative to
a fixed frame and to the body frame, respectively. Owing to the
transformation property inhering in $F_{Qi}^{L\Pi \lambda }$ , associated
with (2) to (4) there are three sets of homogeneous linear equations that
the $F_{Qi}^{L\Pi \lambda }$ have to obey at ETH\ configurations. They read 
\[
\sum_{i^{\prime }}\Pi e^{-i\frac \pi 2Q}G_{i,i^{\prime }}^\lambda
(p(1423))F_{Qi^{\prime }}^{L\Pi \lambda }(ETH)=F_{Qi}^{L\Pi \lambda }(ETH)%
\hspace{1.0in}(7a) 
\]
\[
\sum_{i^{\prime }}\Pi (-1)^LG_{i,i^{\prime }}^\lambda
(p_{12})F_{-Q,i^{\prime }}^{L\Pi \lambda }(ETH)=F_{Qi}^{L\Pi \lambda }(ETH)%
\hspace{1.0in}(7b) 
\]
\[
\sum_{Q^{\prime }i^{\prime }}B_{QQ^{\prime }}G_{i,i^{\prime }}^\lambda
(p(243))F_{Q^{\prime }i^{\prime }}^{L\Pi \lambda }(ETH)=F_{Qi}^{L\Pi \lambda
}(ETH)\hspace{1.0in}(7c) 
\]
where the $(ETH)$ denotes that the coordinates of the four electrons form an
ETH, $G_{i,i^{\prime }}^\lambda $ are the matrix elements of the $\lambda -$%
representation known from the textbook of group theory, and 
\begin{eqnarray*}
\hspace {2.5cm}B_{QQ^{\prime }} &=&<Q^{\prime }|R_{120}^{\stackrel{\wedge }{%
r_1}}|Q> \\
\hspace {2.5cm} &=&\sum_{Q"}D_{Q"Q^{\prime }}^L(0,\theta _0,0)e^{-i\frac{%
2\pi }3Q"}D_{Q"Q}^L(0,\theta _0,0)\hspace {2.8cm}(8)
\end{eqnarray*}
where $\theta _0=\arccos (\sqrt{\frac 13})$ (cf. Fig.1).

The three sets of equations as a whole depend on and only on L, $\Pi ,$ and $%
\lambda $(or S). If there is one (or more than one) set(s) of non-zero $%
F_{Qi}^{L\Pi \lambda }(ETH)$ (not all of them are zero) fulfilling all the
three sets of equations, then we say that $\Psi $ can access the ETH.
Otherwise, $\Psi $ is zero at any ETH configuration disregarding the size
and orientation of the ETH. In this case an INS emerges, and the ETH can not
be accessed. At an ETH, when $\Psi $ is non-zero, from (5) and (6) it can in
general be written as

$\Psi (ETH)=C\stackunder{Qi}{\sum }b_{Qi}^{L\Pi \lambda }D_{QM}^L(-\gamma
,-\beta ,-\alpha )\chi _i^{\stackrel{\sim }{\lambda }}$\hspace{1.0in}(9)

Where the coefficient $b_{Qi}^{L\Pi \lambda }$ is proportional to $%
F_{Qi}^{L\Pi \lambda }(ETH).$ Their values are obtained by solving the
equations (7a) to (7c), thus depend on L,$\Pi $, and $\lambda $ but
absolutely not affected by dynamics, therefore they are model-independent.
Only the coefficient $C$ as a common constant is left to be determined by
dynamics. This fact reveals the decisive effect of symmetry on the
eigenwavefunctions. Evidently, our discussion is quite global and not only
confined in atomic systems. Owing to (7c), different Q-components are mixed
up in (9). Therefore, for the states dominated by ETH-structure (the
wavefunction is mainly distributed around an ETH), Q is far from an
approximately conserved quantum number.

It is noted that when an ETH\ is deformed, the geometric symmetry becomes
weaker, accordingly the constraints becomes fewer. For example, when an ETH
is prolonged along one of its 2-fold axis, say, along k', the prolonged-ETH
is invariant to $O_a$ and $O_b$ , but not $O_{c.}$ Thus, instead of three
sets of equations, the wavefunction at a prolonged-ETH fulfills only (7.a)
and (7.b), but not (7.c). For some cases of L,$\Pi $, and $\lambda $ , it
was found that not only the common non-zero solutions of (7a) to (7c) do not
exist, but also the common non-zero solutions of only (7a) and (7b) do not
exist. In this case the INS located at the ETH\ will extend to the
prolonged-ETH. Since an ETH\ has many ways to deform, the INS located at the
ETH\ has many possibilities to extend. This fact implies that the ETH may be
a source where the INS emerges and extend to its neighborhood. In other
words, in a broad region surrounding the ETH, specific inherent nodal
structures may pour out from the source at the ETH. Accordingly, the
wavefunctions in this broad domain would be seriously affected. Nonetheless,
once a wavefunction can access the ETH, it definitely can access the
neighborhood of the ETH, therefore an ETH-accessible wavefunction is
inherent-nodeless in the broad domain surrounding the ETH.

Another possible source of INS is located at the coplanar squares ( the four
electrons form a square and are coplanar with the nucleus). Let k' be normal
to the plane of the square, let electrons 1 and 2 be located at the two ends
of a diagonal, and let i' be parallel to $\stackrel{\rightarrow }{r_2}-$ $%
\stackrel{\rightarrow }{r_1}$ as before. Then the coplanar square is
invariant to 
\[
O_a^{\prime }=PR_{180}^{k^{\prime }}\hspace{1.0in}(10a) 
\]
\[
O_b^{\prime }=p(1324)R_{90}^{k^{\prime }}\hspace{1.0in}(10b) 
\]
\[
O_c^{\prime }=p_{34}R_{180}^{i^{\prime }}\hspace{1.0in}(10c) 
\]
Thus the $F_{Qi}^{L\Pi \lambda }$ are also constrained at the coplanar
squares by three sets of equations, thereby the accessibility of the
coplanar squares can be identified. As before, the accessibility depends on
L,$\Pi ,$ and $\lambda $ (or S). It is noted that the rotation operators in
eq.(10) do not mix up the $F_{Qi}^{L\Pi \lambda }$ and $F_{Q^{\prime
}i}^{L\Pi \lambda }$with $|Q|\neq |Q^{\prime }|$, but only those with the
same \TEXTsymbol{\vert}Q\TEXTsymbol{\vert} (refer to (10c)). Therefore, at a
coplanar square, an eigenwavefunction can be in general written as

$\Psi (coplanar-square)=\stackunder{|Q|}{\sum }\Psi _{|Q|},$

$\Psi _{|Q|}=C_{|Q|}\stackunder{i}{\sum }[D_{QM}^L(-\gamma ,-\beta ,-\alpha
)+d_{|Q|,i}^{L\Pi \lambda }D_{-Q,M}^L(-\gamma ,-\beta ,-\alpha )]\chi _i^{%
\stackrel{\sim }{\lambda }}$\hspace{1.0in}(11)

For a given \TEXTsymbol{\vert}Q\TEXTsymbol{\vert}, non-zero solutions of
eq.(10) may not exist; in this case we have $\Psi _{|Q|}=0$. Alternatively,
for a given \TEXTsymbol{\vert}Q\TEXTsymbol{\vert}, if a non-zero solution
exists, the eq.(10) as homogeneous equations can determine only the $%
d_{|Q|,i}^{L\Pi \lambda }$ , but not the $C_{|Q|}.$ When more than one
non-zero $\Psi _{|Q|}$ are contained in $\Psi (coplanar-square),$ the $%
C_{|Q|}$ of them are all determined by dynamics. In this case, these
non-zero coefficients $C_{|Q|}$ may be optimized to lower the energy.

For the intrashell states (in these states the 4-valence-electron stay in
the same shell), obviously if the electrons form an ETH the Coulomb
repulsion will be minimized. The other favourable configuration is just the
coplanar square (in some cases the coplanar square is even better than the
ETH, because the moment of inertia of a coplanar square may be larger
resulting in having a smaller collective rotation energy E$_{rot}$). In
general, the wavefunctions of all the low-lying states are mainly
distributed in the domains where the total potential energy is relatively
lower. Therefore, the domains surrounding the ETH and the coplanar squares
are very important to the low-lying states, and the accessibility of these
two domains will be crucial to the low-lying spectrum. Since the search of
solutions of homogeneous linear equations is trivial, we shall neglect the
details but give directly the accessibility in Table 1.

From Table 1 the $^{2S+1}L^\Pi $ states can be classified into three types
as listed in the first column of Table 2. The first type can access both the
ETH and the coplanar squares, these states are essentially inherent
nodeless. The second type contains an INS located either at the ETH or at
the coplanar squares. The third type contains INS both at the ETH and at the
coplanar squares. Let an eigenenergy be divided as a sum of the internal
energy E$_{int}$ and the collective rotation energy E$_{rot}$. Let the
lowest state of a $^{2S+1}L^\Pi $ series be called a first-state. Evidently,
the more the nodal surfaces are contained, the higher the energy. As a
first-state, it will contain the nodal surfaces as least as possible. If
they contain any nodal surface, it was found in [6] that they would contain
only the INS. Thus, we predict that the E$_{int}$ of the first-states of the
first type should be the lowest, in these states the wavefunction can be
freely distributed in the most important domains without nodal surfaces, and
therefore can be optimized to lower the energy. On the other hand, the E$%
_{int}$ of the first-states of the third type should be considerably higher
due to containing more nodal surfaces.

From the classification, the low-lying spectrum can be predicted as follows.
For L=0 states, the $^1S^e$ and $^5S^o$ states would be the lowest two
(because they belong to the second type, while the other L=0 states belong
to the third type). For L=1 states, the $^3P^e$ of the first type would be
the lowest, the $^3P^o$ of the second type would be the second lowest. For
L=2 states, the $^1D^e$ and $^3D^o$ of the first type would be the lowest
two. It is noted that the wavefunctions of a Coulombic system are usually
broadly distributed in the coordinate space due to the long-range character
of the Coulomb force. It is also noted that the domain of the ETH and the
domain of the coplanar square are closely connected (e.g., when an ETH is
flattened along one of its 2-fold axis, it will become to a coplanar
square). Therefore the wave functions of the first-state of the first type
are expected to be smoothly distributed in both domains. This presumption
was actually found in [9] (refer to Fig. 6a of [9]).

It is noted that in general the orientation of a coplanar structure
(relative to L) is crucial to E$_{rot}$. When the normal of the coplanar
structure k' is parallel to L, the moment of inertia is larger ,and
therefore E$_{rot}$ is smaller. Consequently, when a coplanar structure
exists in a $F_{Qi}^{L\Pi \lambda }$ component, the larger the $|Q|$, the
smaller the E$_{rot}$. For the $^3D^o$ state, a coplanar square can exist
only in $|Q|$=1 component (cf. Table 1). However, for the $^1D^e$ state, the
coplanar square can exist both in $|Q|=2$ and 0 components. It is noted that
a first-state will do its best to lower the energy. Thus the $^1D^e$
first-state will be dominated by the \TEXTsymbol{\vert}Q\TEXTsymbol{\vert}=2
component to reduce the E$_{rot}$. For this reason, although both the $^1D^e$
and $^3D^o$ would have favourable internal structure, we predict that the $%
^1D^e$ would be lower than the $^3D^o$ due to having a smaller E$_{rot}$.
The second lowest two L=2 states would be the $^1D^o$ and $^5D^e$ of the
second type. For L=3 states, all the $\ ^3F^e$, $^5F^e$ and $^3F^o$ belong
to the first type. However, the coplanar square can exist in $|Q|=3$
component in $^3F^o,$ exists only in $|Q|=2$ component in $^5F^e$ and exists
only in Q=0 component in $^3F^e$. Thus we predict that the $^3F^o$ is the
lowest, the $^5F^e$ is the second lowest, while the $^3F^e$ is the third
lowest. The $^1F^e$ is the fourth lowest due to belonging to the second type.

In particular, since the internal structure of the first-states of the first
type is most favorable, since among all the states of the first type the $%
^3P^e$ is the one having the smallest L (therefore the smallest E$_{rot}$),
thus we predict that the $^3P^e$ is the ground state. The prediction on the
ordering of the first-states with their quantum numbers is given in Table 2.
For the second-states, they are more energetic and therefore can access a
much broader region, including the region with a higher total potential
energy. Besides the INS, they will contain additional nodal surfaces because
they have to be orthogonal to the first-states. We shall not discuss them in
this paper.

From the above analysis, the first-states of the same type would have
similar internal structures, the energy differences between them arise
mainly from E$_{rot}$. Thus, rotation bands would exist in the spectrum,
this is an important feature. The first-states of the first type including
the $^3P^e,^1D^e,^3D^o,^3F^o,^5F^e,^3F^e,\cdot \cdot \cdot \cdot $ would
form a band headed by the ground state $^3P^e$, therefore this band may be
called the ground-band. The internal wavefunctions $F_{Qi}^{L\Pi \lambda }$
of this band are broadly and smoothly distributed around the ETH and the
coplanar squares without nodal surface. The first-states of the second type
will form two bands. One is composed of the coplanar square-accessible but
ETH-inaccessible states including the $^1S^e,^3P^o,^5D^e,^1F^e,\cdot \cdot
\cdot .$ For these states, there is a source of INS located at the ETH.
Therefore their wave functions may prefer coplanar structure to avoid the
INS, they are expected to be distributed smoothly around a coplanar square.
This band may be called the coplanar square-band. The other band is composed
of the ETH-accessible but coplanar square-inaccessible states including the $%
^5S^o$, $^1D^o,\cdot \cdot \cdot \cdot .$ Their wave functions are expected
to be distributed around the ETH. For these states, the INS\ at the coplanar
square will spoil the stability of the ETH. In fact, a specific mode of
internal oscillation induced by the INS was found in [10]. Therefore, this
band may be called the ETH*-band, where the * implies that an internal
oscillation with one node is inhering. There may be other higher bands,
e.g., composed of the first-states of the third type. However, the above
three bands are the most important bands, we shall not discuss the other
possible higher bands in this paper.

At present both the experimental and theoretical results are not sufficient
to check all the above predictions. Nonetheless, the data derived from an
analysis of the experimental optical spectra by Moore [11] are very
valuable, which are listed in Table 3 for the n=2 intrashell states of
different species. Not all the $^{2S+1}L^\Pi $ symmetries are allowed for a
given shell (due to the limitation in the partial wave of individual
electron and due to the Pauli Principle). However, the optical data for all
the allowed n=2 intrashell states are complete in the case of the ion O$%
^{++} $. These data confirm the predictions very well . The ordering of
levels is exactly as predicted without an exception. Furthermore,
experimentally the ground states of all the 4-valence-electron systems are a 
$^3P^e$ state just as predicted without an exception. Nonetheless, since the
existing data are not complete, more data on quadruply excited states (in
particular those for intrashell states with a larger n) are needed to
justify the predictions.

Owing to the difficulty in calculation, the theoretical literatures on the
quadruply excited states are very few. In 1994 Komninos and Nicolaides have
calculated the intrashell quadruply excited $^5S^o$ states of Be atom by
using the multiconfigurational Hartree-Fock method [12]. They found that in
the first-states of the $^5S^o$ symmetry, the angle between the two position
vectors of any two valence electrons tends to 106$^{\circ }$. On the other
hand, the angle is 109.4$^{\circ }$ for an ETH. Thus their finding supports
that the $^5S^o$ states are dominated by the ETH-structure. In a calculation
based on a r-frozen model [10] the ETH-structure of the $^5S^o$ states was
also found. Besides, the expected INS existing at coplanar squares was
confirmed in [10] (refer to Fig.6a of [10]).

In fact, the distribution of a wavefunction in coordinate space is
determined by two factors, namely the total potential energy and the
inherent nodal structure. Since the first factor is easier to be recognized,
the second factor is crucial. In this paper, the decisive effect of the
inherent nodal structure has been demonstrated, a primary classification
scheme has been proposed, thereby the low-lying spectrum can be
systematically understood. Owing to the rapid progress in experimental
techniques, much more data on the quadruply excited states is expected to be
coming. The qualitative knowledge derived in this paper will be helpful to
understand the coming data.

The generalization of the above procedure to any few-body system is straight
forward. In any case, the sources of INS should be identified, a
classification scheme can be established based on the understanding of the
inherent nodal structure. The existence of the inherent nodal structure is a
great marvel of quantum mechanics. Since different systems are constrained
by symmetry in a similar way, similarity should exist and we can understand
them via an unified point of view.

\hspace{1.0in} \hspace{1.0in} \hspace{1.0in}

REFERENCES

1, T.D.Lee and C.N.Yang,Phys.Rev., 104 (1956) 254

2, F.A.Kaempffer, ''Concepts in Quantum Mechanics'', Acaddemic Pres, 1957

3, J.P.Elliott and P.G.Dawber, ''Symmetry in Physics'', Vol.1 and
2,MacMillan Press LTD.,1979

4, W.Greiner and G.E.Brown, ''Symmetry in Quantum Mechanics'',
Springer-Verlag, 1993

5, T.Rosen, ''Symmetry in Science'', Springer-Verlag, 1997

6, C.G.Bao,Few-Body Systems,{\bf 13},41,1992; Chinese Phys. Lett. 14 (1997)
20; Phys. Rev. Lett 79 (1997) 3475.

7, W.Y.Ruan and C.G.Bao, Few-Body Systems 14 (1993) 25

8, C.G.Bao, X.Z.Yang, and C.D.Lin, Phys. Rev. A55 (1997) 4168.

9, C.G.Bao J. Phys. B (At. Mol. Opt. Phys.) 26 (1993) 4671

10, C.G.Bao, Phys.Rev. A47 (1993) 1752;

11, C.E.Moore, Atomic Energy Levels, NBS Circular 467 ,1971

12, Y.Komninos and C.A.Nicolaides. Pyhs. Rev. A50 (1994) 3782

\hspace{1.0in}

\hspace{1.0in}

Table 1 The accessibility of favorable shapes to the $^{2S+1}L^\Pi $ states.
An empty block implies that the corresponding state can access the ETH
shape. The numbers in the blocks are the $|Q|$ of the non-zero $\Psi _{|Q|}$
, in which the coplanar square can be accessed (e.g., for the $^1D^e$, both
the $|Q|=0$ and 2 components are allowed to access the coplanar squares). A
block with a $\times $ implies that the state can not access the
corresponding shape (e.g., the $^3D^e$ can not access both the ETH and the
coplanar square).

\vskip 1mm 
\begin{tabular}{||c|c|c|c|c|c|c||}
\hline\hline
& $^1S^e$ & $^3S^e$ & $^5S^e$ & $^1S^o$ & $^3S^o$ & $^5S^o$ \\ \hline
ETH & $\times $ & $\times $ & $\times $ & $\times $ & $\times $ &  \\ \hline
coplanar square & 0 & $\times $ & $\times $ & $\times $ & $\times $ & $%
\times $ \\ \hline\hline
& $^1P^e$ & $^3P^e$ & $^5P^e$ & $^1P^o$ & $^3P^o$ & $^5P^o$ \\ \hline
ETH & $\times $ &  & $\times $ & $\times $ & $\times $ & $\times $ \\ \hline
coplanar square & $\times $ & 0 & $\times $ & $\times $ & 1 & $\times $ \\ 
\hline\hline
& $^1D^e$ & $^3D^e$ & $^5D^e$ & $^1D^o$ & $^3D^o$ & $^5D^o$ \\ \hline
ETH &  & $\times $ & $\times $ &  &  & $\times $ \\ \hline
coplanar square & 0,2 & $\times $ & 2 & $\times $ & 1 & $\times $ \\ 
\hline\hline
& $^1F^e$ & $^3F^e$ & $^5F^e$ & $^1F^o$ & $^3F^o$ & $^5F^o$ \\ \hline
ETH & $\times $ &  &  & $\times $ &  & $\times $ \\ \hline
coplanar square & 2 & 0 & 2 & $\times $ & 1,3 & $\times $ \\ \hline\hline
\end{tabular}
\vskip 10mm

\hspace{1.0in}

\hspace{1.0in}

Table 2 A classification and the predicted ordering of levels of the
intrashell first-states based on symmetry consideration. For all the states
with the same L, the states in a higher row is anticipated to be lower in
energy.

\begin{tabular}{|c|c|}
\hline
TYPE & $^{2S+1}L^\Pi $ \\ \hline\hline
2 & $^1S^e,^5S^o$ \\ \hline
3 & $^3S^e,^5S^e,^1S^o,^3S^o$ \\ \hline\hline
1 & $^3P^e$ \\ \hline
2 & $^3P^o$ \\ \hline
3 & $^1P^e,^5P^e,^1P^o,^5P^o$ \\ \hline\hline
1 & $^1D^e$ \\ \hline
1 & $^3D^o$ \\ \hline
2 & $^5D^e,^1D^o$ \\ \hline
3 & $^3D^e,^5D^o$ \\ \hline\hline
1 & $^3F^o$ \\ \hline
1 & $^5F^e$ \\ \hline
1 & $^3F^e$ \\ \hline
2 & $^1F^e$ \\ \hline
3 & $^1F^o,^5F^o$ \\ \hline
\end{tabular}

\hspace{1.0in}

\hspace{1.0in}

Table 3 The energies (in cm$^{-1}$) of the n=2 intrashell states of
4-valence-electron atomic systems given by Moore [11]. The energy of the
ground state $^3P^e$ is zero . All the n=2 intrashell states allowed by the
Pauli Principle are listed . The label $^1S^e(2)$ denotes a second-state,
etc.

\vskip 1mm 
\begin{tabular}{|c|c|c|c|c|}
\hline
TYPE & $^{2S+1}L^\Pi $ & C & N$^{+}$ & O$^{++}$ \\ \hline
2 & $^1S^e$ & 21648 & 32687 & 43184 \\ \hline
2 & $^5S^o$ & 33735 & 47168 & 60312 \\ \hline
3 & $^3S^o$ &  & 155130 & 197086 \\ \hline
& $^1S^e(2)$ &  &  & 343302 \\ \hline
1 & $^3P^e$ & 0 & 0 & 0 \\ \hline
2 & $^3P^o$ & 75256 & 109218 & 142382 \\ \hline
3 & $^1P^o$ &  & 166766 & 210458 \\ \hline
& $^3P^e(2)$ &  &  & 283759 \\ \hline
1 & $^1D^e$ & 10194 & 15316 & 20271 \\ \hline
1 & $^3D^o$ & 64089 & 92238 & 120025 \\ \hline
2 & $^1D^o$ &  & 144189 & 187049 \\ \hline
& $^1D^e(2)$ &  &  & 298289 \\ \hline
\end{tabular}

\end{document}